\documentclass[fleqn,twoside]{article}
\usepackage{espcrc2}

\usepackage{graphicx}

\newcommand{\kbar}{$\bar{K}$}

\title{Strange Tribaryons as $\bar{K}$-Mediated Dense Nuclear Systems
}

\author{
Yoshinori Akaishi\address{College of Science and Technology, Nihon University, Funabashi, Chiba 274-8501, Japan,}\address[RIKEN]{DRI, RIKEN, Wako, Saitama 351-0198, Japan}\thanks{E-mail address: akaishi@post.kek.jp}, 
Akinobu Dot$\acute{\rm e}$\address{Institute of Particle and Nuclear Studies, KEK, 
Tsukuba, Ibaraki 305-0801, Japan}\thanks{Supported by JSPS Research Fellowships for Young Scientists. 
E-mail address: dote@post.kek.jp} and 
Toshimitsu Yamazaki\addressmark[RIKEN]\thanks{E-mail address: yamazaki@nucl.phys.s.u-tokyo.ac.jp}}

\begin{document}

\date{\today}

\begin{abstract}
We discuss the implications of recently discovered strange tribaryons in 
$^4$He(stopped-$K^-, p$)S$^0 (3115)$ and $^4$He(stopped-$K^-, n$) S$^1 (3140)$ 
within the framework of deeply bound \kbar~ states formed on shrunk nuclear cores. 
S$^1 (3140)$ corresponds to $T=0$ $ppnK^-$, whereas S$^0 (3115)$ to $T=1$ $pnnK^-$, 
which is an isobaric analog state of $pppK^-$, predicted previously. 
The observed binding energies can be accounted for by including the relativistic effect 
and by invoking a medium-enhanced \kbar $N$ interaction by 15\%. 
We propose various experimental methods to further study these and related bound systems.    
\end{abstract}


\maketitle

\noindent
{\bf 1. Introduction}\\

In a series of publications in recent years \cite{Akaishi:02,Yamazaki:02,Dote:YKIS02,Dote:02,Dote:03,Yamazaki:04}, we have predicted deeply bound narrow \kbar~nuclear states based 
on bare \kbar $N$ interactions, which were derived from empirical data 
(\kbar $N$ scattering and kaonic hydrogen) together with the ansatz that 
$\Lambda_{1405}$ is a bound state of \kbar $N$. The presence of such hitherto 
unknown kaonic nuclear states results from a very attractive \kbar $N$ interaction 
in the $I=0$ channel, which persists to be strong for discrete bound states of finite nuclei, 
and  causes  not only a strong binding of $K^-$ in proton-rich nuclei, 
but also an enormous shrinkage of \kbar-bound nuclei despite the hard nuclear incompressibility. 
Thus, a \kbar~ produces a bound state with a \kbar-mediated ``condensed nucleus", 
which does not exist by itself. For example, $ppK^-$ with a total binding energy 
of $-E_K = 48$ MeV, $ppnK^-$ with $-E_K = 118 $ MeV, and $pppK^-$ with $-E_K = 97$ 
 MeV. The calculated rms distances  in the $ppnK^-$ system are: $R_{NN}$ = 1.50 fm and 
$R_{\bar{K}-N}$ = 1.19 fm, whereas $R_{\bar{K}-N}$ = 1.31 fm in $\Lambda_{1405}$. 
The $NN$ distance in $ppnK^-$ is substantially smaller than the normal inter-nucleon 
distance ($\sim 2.0$ fm), and the average nucleon density, 
$\rho^{\rm ave}_N = 3.1 \times \rho_0$, is much larger than the  normal density, 
$\rho_0 \sim 0.17$ fm$^{-3}$. An observation of such a deeply bound state would not only confirm 
the underlying physics framework, but would also provide profound information on an 
in-medium modification of the $\bar{K}N$ interaction, nuclear compressibility, 
chiral symmetry restoration, kaon condensation regime and possible transition 
to a quark-gluon phase. Here, the predicted values for the separation energies 
and widths will serve as bench marks to examine the above problems.   

An experimental search for $ppnK^-$ in the $^4$He(stopped-$K^-, n$) reaction \cite{Akaishi:02} 
was carried out by experimental group E471 of  Iwasaki {\it et al.} at KEK. 
The first evidence was found in the proton- emission channel  \cite{TSuzuki:04},
\begin{equation}
^4{\rm He(stopped-}K^-, p)[N'NN\bar{K}]^{Q=0}_{T=1}:~{\rm S}^0(3115), \label{eq:K-p}
\end{equation}
\noindent
where a distinct narrow peak appeared at a mass of  $M = 3117 \pm 5$ MeV/$c^2$ with 
a total binding energy (separation energy of $K^-$) of $S_K =  -E_K 
\equiv -(M - M_p - 2\,M_n  - M_{K^-})c^2 =194 \pm 5$ MeV and a width of less than 25 MeV. 
This state, populated in reaction
 (\ref{eq:K-p}), has a unique isospin, $(T,T_3) = (1,-1)$. 
It was an unexpected discovery, since the $T=1$ bound state of $K^-$ on 
a triton ($(ppn)_{T=1/2}$) had been predicted to be shallow and broad \cite{Akaishi:02}. 
On the other hand, an exotic $T=1$ $pppK^-$ state with $S_K = 97$ MeV had been previously 
predicted \cite{Yamazaki:02,Dote:03}, and the observed S$^0$(3115) can be identified 
as its isobaric analog state. 

An indication of another species was observed in the neutron-emission channel \cite{Iwasaki:04}, 
\begin{equation}
^4{\rm He(stopped-}K^-, n)[N'NN\bar{K}]^{Q=1}_{T = 0}:~{\rm S}^1(3140), \label{eq:K-n}
\end{equation}
in which a peak corresponding to a total mass of $M = 3141 \pm 6$ MeV/$c^2$ with 
a total binding energy of $S_K + BE(^3{\rm He}) = -E_K \equiv -(M - 2\,M_p - M_n - M_{K^-})c^2 
= 169 \pm 6$  MeV and a width less than 25 MeV was revealed. 
This can be identified as the originally predicted $T=0~ppnK^-$. 
Surprisingly, the observed total binding energies of both S$^0$(3115) and S$^1$(3140) 
(194 and 169 MeV, respectively) are much larger than the predicted ones 
(97 and 118 MeV, respectively). Furthermore, the former $T=1$ state lies below 
the latter $T=0$ state, contrary to a naive expectation. 

In the present paper we show that these surprising observations 
can be understood within the framework of deeply bound $\bar{K}$ nuclei. \\ 
 
\noindent
{\bf 2. Spin-isospin structure of strange tribaryons}\\

\begin{figure}
\includegraphics[width=0.48\textwidth]{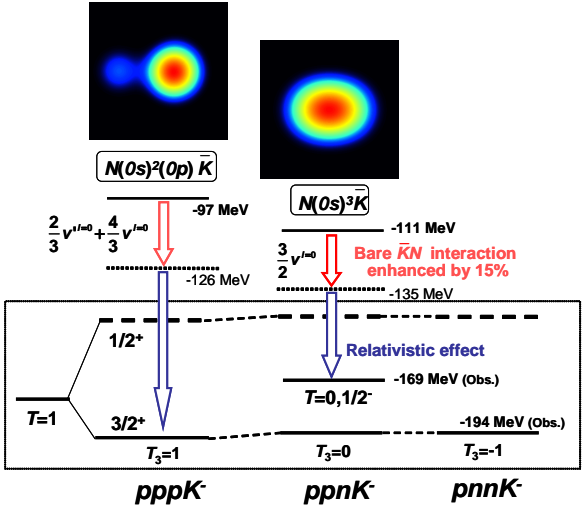}
\vspace{-5mm}
\caption{\label{fig:NNNK-Triplet} Spin-isospin structure of the strange tribaryon 
system $[(NNN)\bar{K}]^{Q}_{(T, T_3)}$. The previously calculated nucleon-density contours 
and energy levels with $E_K$ values are shown on top. The observed S$^0 (3115)$ and S$^1 (3140)$ 
are identified as the $(T,T_3) = (1,-1)$ and $T=0$ states, respectively. }
\end{figure}

Let us first discuss what kinds of states are expected to be low lying 
in the strange tribaryon system. The nomenclature we adopt, 
$[N'NN\bar{K}]^{Q}_{(T,T_3)}$, with $Q$ being a charge and $(T,T_3)$ a total isospin 
and its 3rd component, persists no matter whether the constituent \kbar~ keeps its identity 
or not. We also use a conventional charge-state configuration, such as $ppnK^-$, 
for representing an isospin configuration without any loss of generality. 
The first two nucleons occupy the ground orbital ($0s$), whereas the third nucleon ($N'$) 
in the case of $T_{N'NN} = 3/2$ has to occupy an excited orbital (the lowest one is $0p_{3/2}$). 
An overall view of the tribaryon system together with the experimental information 
is presented in Fig.~\ref{fig:NNNK-Triplet}.

Intuitively speaking, the level ordering depends  on the number of strongly attractive 
$I=0$ \kbar $N$ pairs in each state. Thus,  it is instructive to count the projected number 
of pairs in the \kbar-nucleus interaction in each state.  
The originally predicted $T=0$ state ($ppnK^-$) has partial isospins of  $T_{NN} = 1$ 
and $T_{N'NN} = 1/2$ and spin and parity (including that of $\bar{K}$) of 
$J^{\pi} = 1/2^{-}$, in which the attractive interaction is represented by 
\begin{equation}
V_{N'NN-\bar{K}}^{(T=0)} =  \frac{3}{2} v^{I=0}.
\end{equation}
In this case, the bound ``nucleus" is a shrunk $(ppn)_{T=1/2}$. 
For $T=1$, on the other hand, there are four possible configurations, and the most 
favorite state is a linear combination of two configurations with 
$T_{\bar{K}N'} =1, T_{NN} = 1, T_{N'NN} = 3/2$ and with $T_{\bar{K}N'} =0, T_{NN} = 1, 
T_{N'NN} = 1/2$, in which the attractive interaction is represented by 
\begin{equation}
V_{N'NN-\bar{K}}^{(T=1)} =  \frac{2}{3}v^{I=0'} + \frac{4}{3}v^{I=0}.
\end{equation}
In the case that $v^{I=0'} \sim v^{I=0}$, the attractive interaction for the $T = 1$ state 
amounts to $\sim 2\, v^{I=0}$, which is larger than that for the $T=0$ state. 
This is a key to understanding the level ordering of the $T = 1$ and $T = 0$ states; 
in the former, the stronger $\bar{K}$-core interaction tends to compensate for 
the large internal energy of the $T_{N'NN} = 3/2$ core.  

The predicted $pppK^-$ state corresponds to this lowest $T=1$ state. 
The ``core nucleus" that is bound by $K^-$ is not at all like a triton ($[pnn]_{T = 1/2}$), 
but is close to a non-existing $ppp$. Figure~\ref{fig:NNNK-Triplet} shows an overview of 
the lowest $T=1$ (isospin triplet) and $T=0$ states, together with  the originally predicted 
energy levels and density distributions of $[ppnK^-]_{T=0}$ and $[pppK^-]_{T=1}$ (upper part) 
and the observed energy levels in this framework (lower part). 
Now that $[NNN\bar{K}]^{Q=0}_{(T,T_3) = (1,-1)}$ has been observed as S$^0$(3115), 
another isospin partner S$^{Q=1}_{(T,T_3) = (1,0)}$ should also exist, 
and is expected to appear in a spectrum of $^4$He(stopped-$K^-, n$) with 
a marginal strength \cite{Iwasaki:04}. 

 It should be noted that the larger number  of attractive $\bar{K}N^{I=0}$ in the $T=1$ state 
may cause a lowering of the $T=1$ state, even below the $T=0$ state, although 
the third nucleon in the $T=1$ state should be flipped up to the excited orbital ($0p_{3/2}$). 
In the following we discuss this possibility by addressing the following questions: 
1) the nuclear compression, 2) the relativistic effect, 3) the spin-orbit interaction, 
and 4) the possibility of a medium modification of the bare \kbar $N$ interactions. \\

\vspace{0.5cm}
\noindent
{\bf 3. Relativistic effect on \kbar~binding}\\

The calculations so far made were based on a non-relativistic (NR) treatment of many-body 
systems. For very deeply bound \kbar, however, relativistic corrections are indispensable. 
In this respect it is important to recognize that the \kbar~ bound system is a very peculiar 
one in which the \kbar~ is bound by a non-existing fictitious nucleus, namely, a 
{\it shrunk nuclear core} with a large internal energy (compression energy, 
$\Delta E_{\rm core}$). Thus, to avoid confusion, it is convenient to divide the total 
\kbar~potential (``separating" potential) into the core part and a \kbar-core 
``binding" potential as :  
\begin{equation}
U_K (r) = \Delta E_{\rm core} (r) + U_{\bar{K}{\rm -core}} (r).
\end{equation}
We distinguish between the separation energy of \kbar~ ($S_K$) and the \kbar~ 
binding energy ($B_K \equiv -E_{\bar{K}{\rm -core}}$):
\begin{equation}
-S_K = <\Delta E_{\rm core}> + E_{\bar{K}{\rm -core}},
\end{equation}
where $<\Delta E_{\rm core}>$ is an expectation value of the core compression energy 
with respect to \kbar~ distribution. The calculated shrunk-core energy 
($\Delta E_{\rm core}$ (r)), \kbar-core potentials ($U_{\bar{K}{\rm -core}} (r)$) and 
\kbar-core binding energies ($-E_{\bar{K}-{\rm core}}$) in the $T = 0$ state are shown 
in Fig.~\ref{fig:potential}. 

\begin{figure}
\begin{center}
\vspace{-5mm}
\includegraphics[width=0.3\textwidth]{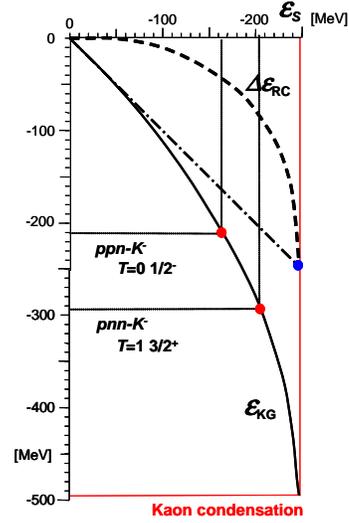}
\end{center}
\vspace{-5mm}
\caption{\label{fig:REL} Relation between the Schr\"{o}dinger energy $(\varepsilon_{\rm S})$ 
and the KG energy $(\varepsilon_{\rm KG})$. $\Delta \varepsilon_{\rm RC} \equiv 
\varepsilon_{\rm KG} - \varepsilon_{\rm S}$ is the relativistic correction.}
\end{figure}

The relativistic effect can be taken into account by using a linearized Klein-Gordon (KG) 
equation for $\bar{K}$, 
\begin{eqnarray}
\left\{ -\frac{\hbar^2}{2\,m_K} \nabla^2 + U_{\bar{K}-{\rm core}} \right\} |\Phi \rangle = 
\left(\varepsilon_{\rm KG} + \frac{\varepsilon_{\rm KG}^2}{2m_K c^2} \right) |\Phi \rangle, \label{KG}
\end{eqnarray}
where $\varepsilon_{\rm KG}$ ($\equiv E_{\bar{K}-{\rm core}}$) is the energy of $\bar{K}$ 
without its rest-mass energy, and $m_K$ the rest mass of \kbar. 
The optical potential, $U_{\bar{K}-{\rm core}}$, is given on the assumption 
that $\bar{K}$ is in a scalar mean-field 
{\it provided by the shrunk nuclear core}. 
When we make a transformation of the KG energy as
\begin{equation}
\left(\varepsilon_{\rm KG} + \frac{\varepsilon_{\rm KG}^2}{2m_K c^2} \right) \longrightarrow 
\varepsilon_{\rm S}. 
\end{equation}
Eq.(\ref{KG}) becomes equivalent to a Schr\"{o}dinger-type equation with an energy solution 
of $\varepsilon_{\rm S}$. Thus, the KG energy can be estimated from a Schr\"{o}dinger solution, 
which we obtain in the NR calculation, by using
\begin{equation}
\varepsilon_{\rm KG} = m_K c^2\, \left( \sqrt{1 + \frac{2\, \varepsilon_{\rm S}}{m_K c^2}}  
-1 \right).
\label{eq:mapping}
\end{equation}
This ``exact" relation means that, when the Schr\"{o}dinger energy $(\varepsilon_{\rm S})$ 
drops down to $-m_K c^2/2$, and the relativistic energy becomes $-m_K c^2$, namely, 
the total mass becomes 0 (``kaon condensation" regime), as shown in Fig.~\ref{fig:REL}. 

\begin{figure}
\begin{center}
\vspace{-15mm}
\includegraphics[width=0.45\textwidth]{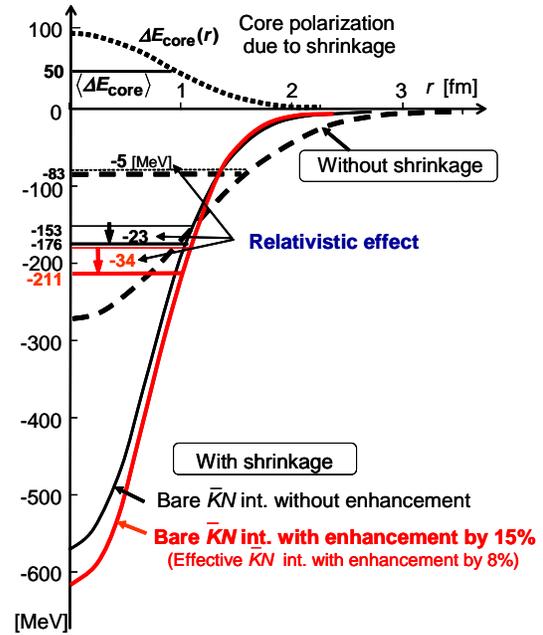}
\vspace{-5mm}
\end{center}
\caption{\label{fig:potential}
Calculated \kbar-core potentials, $U_{\bar{K}-{\rm core}} (r)$, for the $T=0$ $ppnK^-$ state 
in three cases: without shrinkage, with shrinkage with the original bare $\bar{K} N$ interaction 
and with an enhanced bare $\bar{K} N$ interaction by 15\%. The relativistic effects 
on the \kbar-core binding energies ($E_{\bar{K}-{\rm core}}$) are indicated by arrows. 
Also shown is the nuclear core energy, $\Delta E_{\rm core} (r)$, with an average 
value ($<\Delta E_{\rm core}>$).}
\end{figure}

 In this ``mapping" treatment, we also made consistent corrections on 
the threshold energies of $\Lambda + \pi$ and $\Sigma + \pi$ and the complex energy 
of $\Lambda_{1405}$, and obtained re-fitted $\bar{K}N$ interaction parameters. 
It is to be noted that, since the internal energy of the shrunk nucleus is so large, 
the $U_{\bar{K}-{\rm core}}$ for the KG  equation is very deep. Thus, the relativistic treatment 
gives a substantial negative correction to the energy $E_{\bar{K}-{\rm core}}$.

Let us consider the case of $[NNN\bar{K}]_{T=0}$. The original NR total binding energy, 
$E_K = -118$ MeV, was readjusted to $-111$ MeV after taking into account the relativistic 
effect on the $\bar{K} N$ binding in $\Lambda_{1405}$. The nuclear core energy from the core 
shrinkage is $<\Delta E_{\rm core}> \approx 50$ MeV. Thus, we obtain $V_0 = -570$ MeV, 
$W_0 = 18$ MeV and $a_K = 0.923$ fm in the expression for the \kbar-core potential as
\begin{equation}
U_{\bar{K}-{\rm core}} (r)  =  \left(V_0 + i\,W_0\right){\rm exp}\left[-\left(\frac{r}{a_K}\right)^2\right]. \label{eq:K-potential}
\end{equation}
In this case, the relativistic correction is $-23$ MeV, yielding $E_K = -134$ MeV. 
The total binding energy is still smaller by 35 MeV than the experimental value, 169 MeV. 

Next, we consider the case of $[NNN\bar{K}]_{T=1}$. This state has a larger 
$<\Delta E_{\rm core}>$ than the $T=0$ state, because the energy to excite one nucleon 
from the $0s$ shell to the $0p$ is estimated to be 50 MeV. Thus, starting from the NR result, 
we obtain a deeper \kbar-core potential,  $V_0 = -652$ MeV and $W_0 = -12$ MeV, and 
the \kbar~binding energy is subject to a large relativistic correction, 
$\Delta \varepsilon_{\rm RC} = -46$ MeV. 

Roughly speaking, the relativistic effect accounts for about half of the discrepancies in $S_K$. 
It is to be noted that this large correction is a consequence of a  shrinkage of the nuclear 
core. Namely, the large $<\Delta E_{\rm core}>$ (compression energy), which translates 
into a larger negative \kbar~ potential ($U_{\bar{K}-{\rm core}}$), causes a larger 
relativistic correction. However, the resulting total binding energy (143 MeV) is still 
smaller than the observed one (194 MeV).

\vspace{0.5cm}
\noindent
{\bf 4. Spin-orbit splitting}\\

In the $T=1$ state the third nucleon occupies a $0p_{3/2}$ 
orbital and behaves like a compact satellite halo [5], as 
shown in Fig. 1. In our previous prediction we neglected the 
spin-orbit splitting between $0p_{3/2}$ and $0p_{1/2}$. 
Here, we note that the one-body spin-orbit 
interactions may give a large contribution for a shrunk nucleus, 
because it depends on the gradient of the nuclear surface. 
To estimate the effect of the one-body spin-orbit interaction, 
we used the well known Thomas-type 
$\boldmath{\mbox{$l$}} \cdot \boldmath{\mbox{$s$}}$ potential, 
\begin{equation}
V_{ls} (r) = -(\vec{l}\vec{s}) \frac{\hbar^2}{2 M} \frac{2}{r} \frac{{\rm d}}{{\rm d} r} {\rm ln} \left[1 + \frac{ U_{\rm nucl} (r)}{2 M c^2}\right],
\end{equation}and found that $\Delta E (J^\pi=3/2^+) \simeq -5$ MeV and 
$\Delta E (J^\pi=1/2^+) \simeq 10$ MeV in the case of a 
shrunk $0p$ orbital. 

\begin{figure}
\begin{center}
\includegraphics[width=8.0cm]{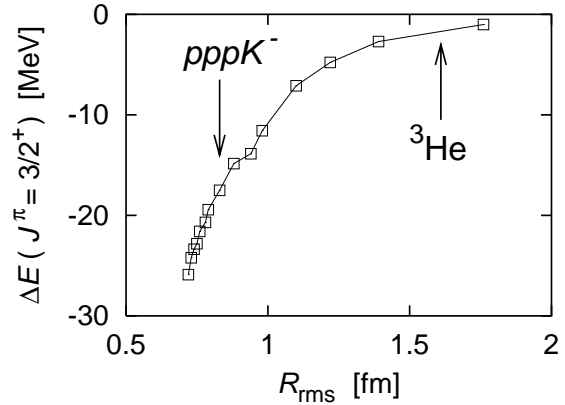}
\end{center}
\vspace{-5mm}
\caption{Spin-orbit energy, $\Delta E(J^{\pi}=3/2^+)$, in $pppK^- (J^\pi=3/2^+, T=1)$ 
 from the $NN$ $LS$
contribution as a function of 
$R_{\rm rms}$,  the root-mean-square radius of the nuclear system. 
\label{NNLS}}
\end{figure}

A further contribution is expected from the nucleon-nucleon two-body spin-orbit 
interaction ($NN$ $LS$ interaction), 
because it is known to be attractive enough to cause the $^3P_2$ pairing in dense 
neutron matter \cite{NNLS_3P2:Tamagaki}.  
We calculated the expectation value of  the sum of 
the $NN$ $LS$ interaction among the nucleons. 
Here, we used the effective 
$LS$ interaction derived from the Tamagaki potential (OPEG) with 
the $g$-matrix method, similarly to our previous studies \cite{Akaishi:02,Dote:YKIS02,Dote:02,Dote:03}. Fig. \ref{NNLS} shows the behavior of the $NN$ $LS$ contribution 
in the $pppK^-$ when the rms radius ($R_{\rm rms}$) is varied.  
The contribution of the $NN$ $LS$ interaction increases
rapidly as the system becomes small. 
The magnitude of $\Delta E (J^\pi=3/2^+)$ is found to increase 
to $\sim 15$ MeV in the shrunk system, whereas 
 it is only $\sim$1 MeV in the normal-size nuclei. 

Thus, the spin-orbit interactions of both kinds make the $T=1$ energy even lower. 
 It is important to find a spin-orbit partner, $J^\pi=1/2^+$, which will give an experimental 
value of the spin-orbit splitting in such a dense nuclear system. From this one can obtain 
information about the size of the shrunk $T=1$ state. According to our calculation, 
the spin-orbit splitting energy is $E(1/2^+) - E(3/2^+) = -3\,\Delta E(3/2^+)\sim 60$ MeV. 

\vspace{0.5cm}
\noindent
{\bf 5. Medium-modified \kbar $N$ interaction}\\

There are still large discrepancies in $S_K$ between theory and observation, even 
after a relativistic correction. They can now be ascribed to a medium-modified bare \kbar $N$ 
interaction that may occur in such a dense nuclear medium. In the present case, 
the average nucleon density, $<\rho(r)>\sim 3 \times \rho_0$, approaches the nucleon compaction 
limit ($\rho_{\rm c} \sim 2.7\,\rho_0$), where a chiral symmetry restoration may occur. 
Similar to the case of the observed pionic bound states \cite{Suzuki:04,Kienle:04}, 
the \kbar $N$ interaction is related to the order parameter of the quark condensate 
and is expected to be enhanced as chiral symmetry is restored. Thus, we tuned the bare 
\kbar $N$ strength by a small factor, and recalculated the total binding energies to find 
the most suitable enhancement parameter. Since this modification causes a change 
in the relativistic correction, we iterated all of these corrections consistently. 
The following enhancement was found to account for both S$^0$(3115) and S$^1$(3140) 
simultaneously: 
\begin{equation}
\frac{\bar{K}N}{\bar{K}N^{\rm bare}} \sim 1.15. \label{eq:enhancement}
\end{equation}
The final results after this tuning are also presented in Fig.~\ref{fig:potential}. 
The \kbar-core potential strength, $U_{\bar{K}-{\rm core}}$, is now $-618 - i\, 11$ MeV 
with $a_K = 0.920$ fm for the $T=0$ state. 

Using the enhanced \kbar$N$ interaction strength and also taking into account 
the relativistic effect, we recalculated the binding energy and width of the most basic 
\kbar~nuclear system, $ppK^-$. The results are $M = 2284$ MeV$/c^2$ ($S_K = 86$ MeV) and 
$\Gamma = 58$ MeV, in contrast to the original non-relativistic values, 
$M = 2322$ MeV$/c^2$ ($S_K = 48$ MeV) and $\Gamma = 61$ MeV. It is important to find 
this state experimentally so as to establish a solid starting gauge for 
more complex \kbar~ bound systems.\\

\vspace{0.5cm}
\noindent
{\bf 6.  Energy difference among the isotriplet states}\\

Although the observed $T=0$ and $T=1$ states support the theoretical expectation 
for nuclear shrinkage, a direct experimental verification, if possible, 
would be vitally important. 
We examine the energy difference of the isobaric analog states of the isotriplet, 
which is related to the strong-interaction mass term and the Coulomb displacement energy as
\begin{equation}
\Delta E_{\rm sum} (T_3)=  \Delta E_{\rm mass} (T_3)+ E_{\rm C} (T_3).
\end{equation}
We estimated the mass term, $\Delta E_{\rm mass}$, as the deviation of the sum of 
the constituent particle masses ($p, n, \bar{K_0}$ and $K^-$) from the central value 
by using the calculated isospin eigenstates in terms of the charge states 
(the $(T, T_3) = (1, 1)$ and $(1, -1)$ isospin eigenstates have mixtures of 
70\% $pppK^-$ + 30\% $ppn\bar{K_0}$ and of 30\% $pnnK^-$ + 70\% $nnn\bar{K_0}$, respectively). 
The values of $\Delta E(T_3)$ are listed in Table~\ref{tab:isotriplet}.

\begin{table}
\caption{\label{tab:isotriplet}  Energy differences in the isotriplet states. 
$\Delta E_{\rm mass}$, constituent rest energy (relative); $E_{\rm Coulomb}$, Coulomb energy; 
$E_{\rm sum}$, total energy difference (relative).  All in MeV. 
}
\begin{tabular}{|l|rrr|}
\hline
$(T,T_3)$          &    $ (1,1) $        & $ (1,0)$             &   $ (1,-1) $        \\
\hline
$Q$                     &   2                     &    1                     &   0                      \\
Charge states     & $pppK^-$        & $ppnK^-$      & $pnnK^-$\\
                            & $ppn\bar{K}^0$ &$pnn\bar{K}^0$ & $nnn\bar{K}^0$ \\ 
\hline
$\Delta E_{\rm mass}$        &  $-2.4$       &  $-0.7$    & 2.4  \\
$E_{\rm C} ({\rm total})$    &   $-0.6$   &  $-1.3$                &   $-0.7$            \\
~~$E_{\rm C} (NN)$    &   3.7           &  1.1               &   0           \\
~~$E_{\rm C} (\bar{K}N)$ &  $-4.3$   &   $-2.4$         &  $-0.7$            \\
$\Delta E_{\rm sum}$                 & $-3.0$ &  $-2.0$   & 1.7     \\     
\hline
\end{tabular}
\end{table}

For  estimating $E_{\rm C}$ we calculated the Coulomb energy of each particle pair 
using the total wavefunction. The results, presented in Table~\ref{tab:isotriplet}, 
show that the Coulomb energies of the $NN$ and $\bar{K} N$ pairs are 0 and $-0.7$ MeV, 
respectively, for the $T_3 = -1$ state, whereas they increase in magnitude to 3.7 and 
$-4.3$ MeV, respectively, in the $T_3 = 1$ state, which are, however, nearly cancelled 
by each other. Thus, the total Coulomb energies for the two isospin states remain nearly zero. 
On the other hand, a naive estimate of the Coulomb energy assuming a uniformly charged sphere 
for a fictitious $pppK^-$ would give $E_{\rm C} (T_3=1) -  E_{\rm C} (T_3=-1) \sim 
(3/5) Q^2e^2/R \sim 2.1$ MeV, if one takes $R_{\rm rms} = 1.61$ fm, the ordinary nuclear 
radius for $A = 3$. The different cases can be distinguished experimentally, 
if the two (or three) isobaric analog states are produced and identified. 
In the next section we propose some experimental methods.\\

\noindent
{\bf 8. Role of $\Lambda_{1405}$ in the S$^0$ population}\\

\begin{figure}
\begin{center}
\includegraphics[height=7cm]{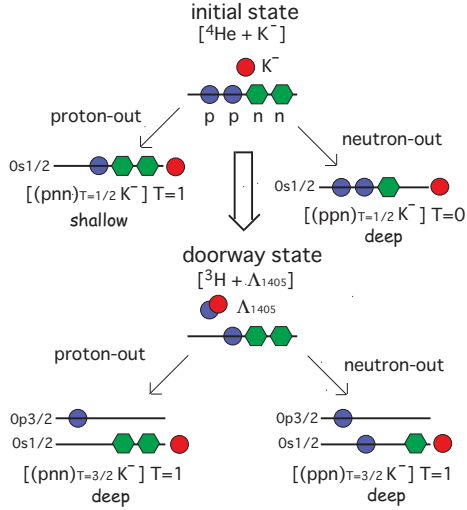}
\vspace{-5mm}
\end{center}
\caption{\label{fig:doorway} Population mechanism of 
the $[N'NN\bar{K}]_{T=1}$
through the $\Lambda_{1405}$ doorway in 
$^4$He(stopped-$K^-,x)$.}
\end{figure}

In the $^4$He(stopped-$K^-, p)$ reaction with Auger-proton emission, 
the three nucleons in the target $^4$He are expected to remain in the $0s$ orbital. 
Then, why can the $T=1$ state with a shrunk core of $T=3/2$ be populated? The key to understand 
this process is the role of $\Lambda_{1405}$ ($\equiv \Lambda^*$) as a doorway; 
the formation of $\Lambda^*$ in the $K^-$ absorption at rest by $^4$He is known to occur 
with a substantial branching ratio \cite{Riley:75}. This doorway state can lead to core 
excited \kbar~ states:
\begin{eqnarray}
K^- + ``p" &\rightarrow& \Lambda^*,\\
   \Lambda^* + ``pnn" &\rightarrow& [(pnn)_{T=3/2}K^-]_{T=1} + p,\\
                                                &\rightarrow& [(ppn)_{T=3/2}K^-]_{T=1} + n,
\end{eqnarray}   
where the proton from $\Lambda^* = pK^-$ falls onto the $0p$ orbital, 
as shown in Fig.~\ref{fig:doorway}. 
Likewise, the doorway $\Lambda^*$ leads to many other \kbar~bound states, 
such as  $\Lambda^* p \rightarrow ppK^-$, as emphasized in \cite{Yamazaki:02,Yamazaki:01}.\\

\noindent
{\bf 9. Future experiments}\\

\begin{figure}
\begin{center}
\vspace{0cm}
\includegraphics[width=0.45\textwidth]{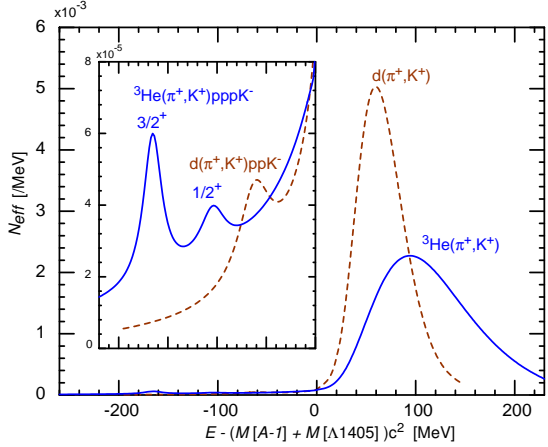}
\caption{
Spectra ($N_{\rm eff}$) of ($\pi^+,K^+$) reactions at $p_{\pi} = 1.5$ GeV/$c$ on $d$ and $^3$He 
as functions of $E_K - [M(A-1) + M(\Lambda_{1405})]c^2$, calculated based on the $\Lambda^*$
 doorway model. Not only the $J^{\pi}=3/2^+$ state, 
but also its spin-orbit partner ($1/2^+$) with a splitting of 60 MeV are incorporated.}
\label{fig:piK}
\end{center}
\end{figure}

A direct reaction to produce $pppK^-$ via $^3$He$(K^-,\pi^-)$ and $^3$He$(\pi^+,K^+)$ was 
proposed in \cite{Yamazaki:02}. Spectral functions (``effective nucleon numbers") in the 
missing mass, as shown in Fig.~\ref{fig:piK}, were calculated based on 
the $\Lambda_{1405}$ doorway model. The spectrum shows the spin-orbit pair 
($J^{\pi} =  3/2^+$ and $1/2^+$) states with a calculated splitting of ~60 MeV. 
Such a pair, if observed, will give important information on the size of the system.

Another way to produce $pppK^-$ is to use a cascade reaction in a light target (say, $^4$He), 
such as
\begin{eqnarray}
p + ``n" &\rightarrow& \Lambda^* + K^0 + p,\\
\Lambda^* + ``ppn" &\rightarrow& pppK^- + n.
\end{eqnarray}
In the second process, the energetic ``doorway particle", $\Lambda_{1405}$, produced 
by an incident proton of sufficiently large kinetic energy, knocks on one of the remaining 
nucleons and/or hits the remaining nucleus to form a kaonic bound system 
(``\kbar-transfer" reactions). $\Lambda^*$ compound processes induced by $(K^-,\pi^-)$ 
and $(\pi^+,K^+)$ may also produce kaonic systems. Recently, it has been pointed out 
that a fireball in heavy-ion collisions can be a source for kaonic systems \cite{Yamazaki:04}. 
In all of these reactions one can identify a \kbar~cluster by invariant-mass spectroscopy 
following its decay, such as 
\begin{equation}
pppK^- \rightarrow p + p + \Lambda.
\end{equation}

\noindent
{\bf 10. Concluding remarks}\\

In the present paper we have shown that the observed strange tribaryons, S$^0$(3115) and 
S$^1$(3140), can be understood as the $(T, T_3)=(1,-1)$ and $T=0$ $N'NN\bar{K}$ bound states 
with shrunk nuclear cores of $T=3/2, J=3/2$ and $T=1/2, J=1/2$, respectively. 
The fact that S$^0$(3115) lies below S$^1$(3140) strongly supports the prediction 
that the three nucleons are in a ``non-existing nucleus" $(N'NN)_{T=3/2,J=3/2}$ with which 
the attractive $I=0$ \kbar $N$ attraction is maximal. The spin-orbit splitting, enhanced 
in a condensed nucleus, helps to further lower the $T=1$ state. 
The  observed binding energies, which are substantially larger than the predicted 
non-relativistic values, are partially accounted for by the relativistic effect on 
the $\bar{K}$ and partially by invoking an enhanced bare $\bar{K} N$ interaction. 
The enhancement may indicate a partial restoration of chiral symmetry and/or a transition 
to a 11-quark-gluon phase. The observed deep \kbar~binding indicates that the system 
is approaching the kaon condensation regime \cite{Kaplan:86,Brown:94}. 
These discoveries have demonstrated that narrow deeply bound states of \kbar~exist, 
as we have predicted, in contrast to the prevailing belief and claim for a shallow 
\kbar~ potential \cite{Ramos:00,Cieply:01}, which may apply to unbound continuum systems.  \\

\noindent   
{\bf Acknowledgement}\\

 We would like to thank Professor M. Iwasaki and his group for their communication  concerning  the experimental results. The present work is supported by Grant-in-Aid of Monbukagakusho of Japan.

\end{document}